\documentclass[10pt, conference,letterpaper]{IEEEtran}

\ifCLASSINFOpdf
\else
\fi
\usepackage{subfigure}
\usepackage{graphics}
\usepackage{epstopdf}
\usepackage{epsfig}
\usepackage{color}
\usepackage{url}
\usepackage{algorithmic}
\usepackage{algorithm}
\usepackage{microtype}
\usepackage{cite}
\usepackage{placeins}

\definecolor{forestgreen}{RGB}{34,139,34}

\begin{document}
%
\title{SybilControl: Practical Sybil Defense with Computational Puzzles}


\author{\IEEEauthorblockN{Frank Li}
\IEEEauthorblockA{Department of Computer Science\\
Massachusetts Institute of Technology\\
Cambridge, MA\\
frankli@mit.edu}
\and
\IEEEauthorblockN{Prateek Mittal$^1$, Matthew Caesar$^2$, Nikita Borisov$^1$}
\IEEEauthorblockA{$^1$Department of Electrical and Computer Engineering\\
 $^2$Department of Computer Science\\
University of Illinois at Urbana-Champaign\\
Urbana, IL\\
\{mittal2, caesar, nikita\}@illinois.edu}
}


%


\maketitle

\begin{abstract}
 
Many distributed systems are subject to the {\em Sybil attack}, where an
adversary subverts system operation by emulating behavior of multiple distinct
nodes.  Most recent work to address this problem leverages social networks to
establish trust relationships between users.  However, the use of social
networks is not appropriate in all systems, as they can be subverted by social engineering techniques, 
require nodes in a P2P network to maintain and be aware of social network information, and may
require overly optimistic assumptions about the fast-mixing nature of social
links.

This paper explores an alternate approach. We present SybilControl, a novel,
decentralized scheme for controlling the extent of Sybil attacks.  SybilControl is an admission control mechanism
for nodes in a distributed system that requires them to periodically 
solve computational puzzles. SybilControl consists of a distributed
protocol to allow nodes to collectively verify the computational work of other nodes, and
mechanisms to prevent the malicious influence of misbehaving nodes that do not perform the computational work.  
We investigate the practical issues involved with deploying
SybilControl into existing DHTs, particularly with resilient lookup protocols. 
We evaluate SybilControl through simulations and find that SybilControl retains low
overhead and latency.  Additionally, even when the adversary controls 20$\%$ of the system's computational resources, 
SybilControl-enabled DHTs can be configured to maintain lookup performance at over 99$\%$ success rate using low communication overhead.

\end{abstract}


%
\IEEEpeerreviewmaketitle

\section{Introduction}
 
Decentralized distributed systems, such as peer-to-peer networks, are
particularly susceptible to the {\em Sybil attack} \cite{douceur}. In this attack, an
adversary exploits the low cost-of-entry into a system by obtaining
multiple identities, thus taking the guise of many distinct nodes. These {\em Sybil nodes} can then collude to launch further attacks, for example by taking over resources and disrupting connectivity, to subvert the system's operation.
Researchers have documented this vulnerability in real-world systems, including
the Maze P2P file-sharing system \cite{maze-collusion, maze-freerider} and
Vanish \cite{vanish, defeat-vanish}, a system for self-destructing data that
relies on a distributed hash table (DHT).

To address this problem, recent research has focused on leveraging information from
social networks to protect honest users, resulting in several state-of-the-art
decentralized approaches\cite{sybillimit,sybilguard,sybilinfer,whanau:nsdi10}.
The key observation behind these works is that it is expensive for a malicious
adversary to establish social links with honest users, and hence social network
graphs can be used to detect and mitigate Sybil attacks.  

However, the reliance of these techniques on social networks has disadvantages. The use of a social network is not 
appropriate in many systems, as social relationships require P2P nodes to be aware of their social contacts and users to exert manual effort to
set up and maintain accurate relationships. Also,
social network-based approaches operate on the assumption that the number of connections between
Sybil nodes and honest nodes in the network is small relative to the number of
honest nodes. 
However, recent studies \cite{all-your-contacts, analysis-social, context-spam} indicate social networks are prone to social engineering
attacks (for example, some Facebook users agree to all friend requests regardless of whether
the other party is known to them). 
Furthermore, these defenses often require the social networks to be
\textit{fast-mixing}, meaning a random walk on the graph approaches the
stationary distribution relatively quickly. However, recent investigations~\cite{fast-mixing} into the mixing times of real-world social networks have
found that they may not be as fast as what the approaches
assumed. These disadvantages lead to questions about the practicality of social
network methods in real-world contexts. 

To address this, we propose SybilControl, a novel, decentralized Sybil defense scheme. 
Inspired by the conceptual work of Borisov~\cite{borisov}, SybilControl goes beyond the theory and is the
first practical, decentralized scheme to mitigate Sybil attacks
that does not require the use of social networks.
It is based on the insight that if
an adversary with finite resources must dedicate some resources to support each node in a system,
then it can only support a limited number of adversarial nodes. SybilControl enforces that
nodes dedicate computational resources to support their system identities.
This enforcement is conducted through two main components: a distributed
protocol to allow nodes to collectively verify computational work of other nodes,
 and defense mechanisms that protect honest nodes from the influences of
misbehaving nodes that do not perform computational work.

In particular, a SybilControl-enabled node
can distribute challenges to its neighbors, and can verify that those
neighbors recently used the challenges to solve computational puzzles, which require significant
amounts of computation to solve but are easy to verify. Challenges are propagated throughout the network in a scalable fashion using aggregation, enabling nodes
to verify other nodes that are not their adjacent neighbors.
To provide the property that the number of identifiers owned by a user
remains commensurate with the computational power of the user,
SybilControl nodes periodically re-issue fresh challenges, forcing other nodes
to periodically solve new puzzles to remain in the system. Defense mechanisms protect
honest nodes from Sybils that do not perform computational work, preventing those Sybils from subverting the system.
To reduce overheads, SybilControl can operate over sessions already maintained
by the underlying system to be secured (e.g., finger and successor relationships in DHTs).

While SybilControl can be applied to a wide variety of distributed
systems, to grapple with practical issues, we deploy it in the context
of the Chord DHT~\cite{chord}.
We evaluate a SybilControl-enabled instance of Chord through simulations, and find that it introduces minimal communication
overheads while retaining the low-latency forwarding properties of Chord.
Additionally, it can remain robust even when the adversary controls 20$\%$ of the system's computational resources, maintaining a lookup success rate above 99$\%$ with low communication overhead.

The remainder of the paper is structured as follows: In
Section~\ref{sec:relwk}, we discuss related works in Sybil defense. In
Section~\ref{sec:overview}, we provide an overview of our approach, followed by the design
details of SybilControl in Section~\ref{sec:design}. Section \ref{sec:incorp} documents the deployment of SybilControl, specifically for DHTs.
The performance and effectiveness of SybilControl is evaluated in Section~\ref{sec:eval}, followed by concluding remarks in
Section~\ref{sec:conclusion}.

\section{Related Work}
\label{sec:relwk}

After the initial classification of the Sybil attack \cite{douceur}, several
centralized Sybil defense schemes \cite{secure-p2p-routing, comp-tree} were
proposed that focused on using a central authority to certify nodes for
admission control and limit the number of Sybil identities. However, the
centralized nature of these schemes conflict with many of the guiding
principles of distributed systems, and a central authority creates trust
issues, a single point of failure, and possible performance bottlenecks.

In an alternate direction, several proposals have focused on
decentralized schemes that work better in conjunction with distributed systems.
One approach \cite{bootstrap-graph} is to utilize the bootstrap graph of DHTs.
The idea is that a small number of malicious users would introduce, or
bootstrap, many Sybil nodes into the system, so the bootstrap graph for these
nodes would be highly connected. However, it is possible that honest nodes
might also bootstrap many other honest nodes, making it difficult to accurately
distinguish honest from malicious users. A distributed registration system was
introduced in \cite{dinger} that regulated the number of identities a specific
IP address could obtain. However, it is possible for adversaries to control
multiple IP addresses through spoofing or prefix hijacking attacks. 
Additionally, multiple honest users could share the same IP address if they are
behind a NAT, which makes this approach less practical.

The most recent schemes have focused on exploiting social network information
\cite{sybillimit,sybilguard,sybilinfer,whanau:nsdi10}. SybilGuard,  its
successor SybilLimit, and Whanau \cite{sybillimit,sybilguard, whanau:nsdi10}
provide protocols to defend against Sybil attacks, based on random walks over a
social network. SybilInfer \cite{sybilinfer} uses Bayesian inference based on
information gathered from the social network to distinguish between malicious
and honest nodes. However, all of these defenses assume that social networks
are ``fast-mixing'', which a recent study \cite{fast-mixing} has shown is not
true in real-world networks to the extent these schemes assume. This leaves
open questions about the strengths of the security guarantees these mechanisms
provide and their true real-world effectiveness. Furthermore, use of social
networks may not be appropriate in all systems since nodes must then be aware of their social contacts. Also, social relationships require manual effort on the part of human users to instrument and correctly maintain.

The concept of using computational puzzles for Sybil defense is not unique to
SybilControl. A related approach by Borisov \cite{borisov} also discussed
challenge distribution and puzzle solving verification in overlay networks, and
inspired several key techniques at the core of our scheme. However, the work
provided only a conceptual description, without any implementation, analysis or
evaluation on effectiveness. It did not investigate necessary modifications to
existing distributed system protocols to maintain system performance under the
scheme. The scheme also had several flaws in its design,
and was incomplete in that it did not handle misbehaving nodes that do not
solve puzzles. Furthermore, node capacity heterogeneity, a common argument against the
practical use of computational puzzle, was not dealt with. SybilControl
addresses all these short-comings.

\section{Overview}
\label{sec:overview}

The goal of SybilControl (or other Sybil prevention schemes) is not to completely prevent adversaries from joining the
system, but rather to place a limit on the number of additional Sybil nodes adversaries can join, thereby
preventing them from obtaining significant influence over the system.
SybilControl operates towards this goal using the following insight: if a computational cost is incurred
by nodes before they are allowed to join the system, then adversaries with
finite resources will have an upper bound on the rate they can acquire identities.
Moreover, if nodes are required to periodically repay this computational cost to
retain their identifiers, then the number of identifiers that can be maintained
by the adversary will be limited. 

To leverage this insight, SybilControl
controls admission and retainment of nodes
into a system. To do this, SybilControl provides a distributed enforcement
mechanism to allow network participants to collectively verify that their
neighbors are paying computational costs (through the use of puzzles) to remain
in the system. In particular, SybilControl provides mechanisms to 
address two key challenges:
\begin{figure}[t!]
        \center{\includegraphics[width=0.32\textwidth, height=0.2\textheight]
        {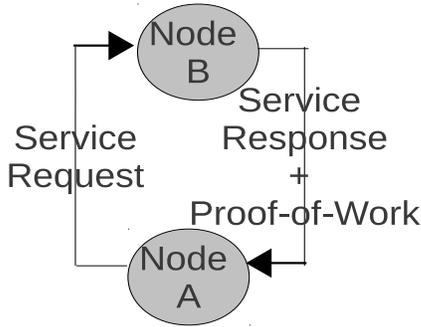}}
       
        \caption{\label{fig:reqService} SybilControl-enabled nodes require proof-of-work when accepting services.}  
 \end{figure}

\subsection{Collectively verifying a node:}
In a distributed system, we lack a centralized authority to verify puzzle solutions of new
arrivals. To address this, SybilControl allows decentralized groups of nodes to collectively
verify the computational work done by their neighbors.
For example in Figure \ref{fig:reqService}, before a
node $A$ trusts communication with another node $B$, $A$ requires $B$ to prove that it
recently solved a puzzle. If $B$ is a malicious Sybil node and chooses not to
solve the puzzles, it will not be able to provide proof-of-work. Defense
mechanisms in SybilControl protect honest nodes like $A$ from using $B$,
essentially making $B$ non-functional in the system and preventing it from
doing harm. This forces adversaries to use only puzzle-solving Sybils, of which they can support a limited number.

\begin{figure*}[t]
        \centering
        \subfigure[]{\includegraphics[width=0.32\textwidth]{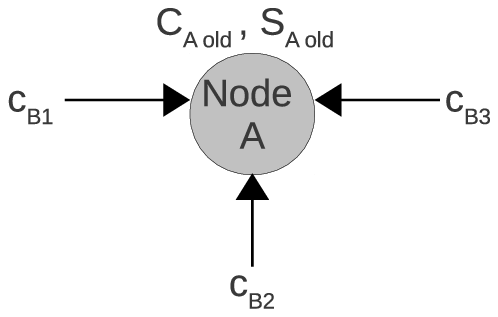} \label{fig:challengeDistrib}}
        \subfigure[]{\includegraphics[width=0.32\textwidth]{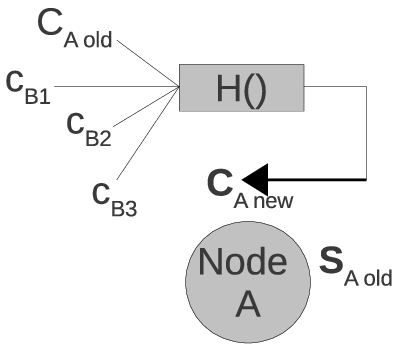} \label{fig:newChallenge}}   
        \subfigure[]{\includegraphics[width=0.32\textwidth]{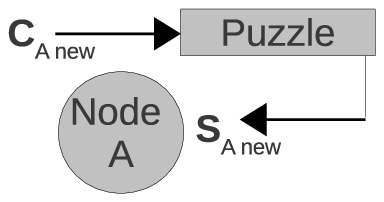} \label{fig:puzzleSolve}}     
         \caption{\label{fig:overview} Overview of SybilControl.  (a) Node A maintains a puzzle solution $s_{A-old}$ and its associated challenge $c_{A-old}$. Node A receives challenges $c$ from other nodes $B_1$, $B_2$, and $B_3$, which (b) are used along with node A's old challenge to periodically compute a new challenge $c_{A-new}$. (c) Node A periodically generates a fresh puzzle solution based on its new challenge.}  
 \end{figure*}

To establish recurring proof-of-work, SybilControl uses a distributed {\em
collective verification scheme}, where nodes periodically challenge each other to solve new puzzles. 
Following this scheme, if a group of nodes $B_1$, $B_2$, and $B_3$ collectively
desire to communicate with another node $A$, they each periodically
create, record, and send new challenges to $A$, as in Figure \ref{fig:challengeDistrib}. $A$
also periodically creates a new challenge using the latest received challenges, as in Figure \ref{fig:newChallenge}, and uses
that new challenge to solve a new puzzle, as shown in Figure \ref{fig:puzzleSolve}.
When any one of the nodes requests a service from $A$, for example node $B_1$, $A$ responds with the service as well as
information from the latest puzzle it solved. If $B_1$ still has recorded the
original challenge used in the puzzle, $B_1$ can verify $A$'s puzzle solution.
The duration for which $B_1$ records challenges can put a 
bound on how recent $A$'s solution is, validating its timeliness. If all nodes in the network
formed a single group and collectively challenged each other,
then all pairs of nodes can perform direct verification.

\subsection{Verifying across multiple hops:}
Performing direct verification between all pairs of nodes in the network can be prohibitively expensive.
While it may scale to systems that communicate in a full mesh (e.g., one-hop overlay networks),
many systems restrict the number of neighbors a node is allowed to communicate with for scaling
purposes (e.g., in DHTs like Chord, nodes maintain regular communication relationships with only a 
logarithmic number of adjacent neighbors). To support these systems, SybilControl
provides a {\em multi-hop verification scheme}.

To do this, nodes in SybilControl only exchange challenges with their
neighbors.  Neighbor relationships may be selected arbitrarily, or in a manner
based on the instrumented distributed system (e.g., when applying SybilControl
to the Chord DHT, challenges may be exchanged only with a node's fingers and
successors).  Each node, as in Figure \ref{fig:newChallenge}, then performs a {\em cryptographic aggregation} step to
combine the challenges received by its neighbors, and uses this aggregation as
input to construct its own challenges to be sent to its neighbors. This process
repeats, allowing a node's challenge to be distributed throughout the network.
SybilControl provides a multi-hop verification mechanism, allowing a node to
check whether its challenge is indirectly incorporated by a remote node.  This allows
indirect proof-of-work to be established between a node and non-neighbors.

\section{SybilControl Design}
\label{sec:design}

In this section, we discuss the details of SybilControl's design. SybilControl limits the extent of Sybil attacks by enforcing that nodes in the system perform periodic computational work to remain in the system. SybilControl achieves this using two parts: a collective verification scheme and protection mechanisms. The collective verification scheme establishes an environment where nodes can collectively verify the computational work of other nodes. The protection mechanisms utilize this verification environment to allow honest nodes to protect themselves from Sybil nodes. These mechanisms eliminate the threat of Sybil nodes that do not solve puzzles. This forces adversaries to use only verifiable Sybils, of which they can only support a limited number. Section \ref{sec:collect} describes the collective verification scheme, while Section \ref{sec:protect} details the protection mechanisms.

\subsection{Collective Verification Scheme}
\label{sec:collect}

The collective verification scheme of SybilControl allows nodes to challenge one another to solve computational puzzles and to collectively verify the completion of the work. This creates an environment where honest nodes can detect and avoid contact with misbehaving nodes that do not perform work. Section \ref{sec:distrib} discusses the distribution of puzzle challenges, while Section \ref{sec:puzVer} illustrates the process of solving puzzles and directly verifying neighbors. Multi-hop verification is used to verify non-neighbor nodes, as described in Section \ref{sec:pathVer}.

\subsubsection{Distributing Challenges}
\label{sec:distrib}

SybilControl-enabled nodes must efficiently distribute numerical challenges to other nodes that are used in solving puzzles. Since in most distributed systems, nodes regularly ping their neighbors to ensure availability, these ping messages provide a convenient medium for carrying challenges. We use a modified ping message that additionally includes the challenge. This results in challenge distribution only directly to neighbors, an efficient method which scales with the underlying system and requires little overhead.

The difficulty with this method is propagating a challenge throughout the network while communicating with only neighbors. To overcome this, a node's new challenge is computed periodically based off a hash of the node's latest distributed challenge, the latest received challenges, and the identifiers for the nodes that sent those challenges, as in Figure \ref{fig:newChallenge}. Let node $A$ be the neighbor of $m$ nodes $B_1$,...,$B_m$. If the latest challenges $A$ has received from those nodes are $c_{B_1}$,...,$c_{B_m}$, then $A$ computes a new challenge of: 
\begin{center}
$c_{A_{new}}$=$H(B_1 || c_{B_1}||...||B_m||c_{B_m}||c_{A_{old}})$ 
\end{center}
where $||$ represents concatenation and $H()$ can be a secure hash function such as SHA-2 \cite{sha}. Since new challenges are constructed using a cryptographic aggregation of received challenges, challenges are indirectly propagated throughout the network without direct all-to-all distribution. When $A$ distributes $c_{A_{new}}$, $A$'s neighbors will receive a challenge that indirectly includes the challenges of $B_1$,...,$B_m$.

Each time a new challenge is computed, the data associated with its construction must be maintained for a period of time. This allows future verification of puzzles that use the new challenge. The newly computed challenge and the previously computed challenge are stored, along with a table of the associated received challenges, referred to as the \textit{state records}. So when $c_{A_{new}}$ is computed, it is recorded with $c_{A_{old}}$ and the state record mapping $B_i$ to $c_{B_i}$ for $i$ in $[1,m]$. 

The duration for which this data is maintained depends on the length of time it takes a challenge to propagate a network and be a viable candidate for puzzle solving. If each node sends out pings every $p$ seconds and the network has diameter $d$, then assuming $p$ seconds is longer than network latency, it takes at most $pd$ seconds for a challenge to indirectly propagate across the network. Note that in efficient overlay networks, $d=O(lg N)$. Additionally, if a node $A$ just received a challenge from node $B$, it will receive another challenge from $B$ within $2p$: at most $p$ seconds before $B$ sends a new ping and at most $p$ seconds before the ping message arrives. So if $A$ doesn't solve a puzzle within $2p$, the challenge it just received from $B$ will not be used to solve a puzzle. Using this insight, the maximum time it could take a challenge used for a puzzle to propagate the diameter of a network is $p(d+2)$. If a puzzle is solved every $s$ seconds, a node's puzzle could be based on a specific challenge for a total of $2s$: $s$ seconds for solving the puzzle with that challenge and another $s$ seconds before a new puzzle is solved, replacing the need to record that old challenge. Therefore, records should be maintained for $2s+p(d+2)$ seconds.

\subsubsection{Solving Puzzles and Directly Verifying Neighbors}
\label{sec:puzVer}

SybilControl-enabled nodes must solve puzzles periodically. A node uses the latest computed challenge to solve the puzzles, which implies the puzzle solutions are dependent on received challenges. Ideally, the challenge used to solve a new puzzle should be fresh, meaning it was computed from newly received (directly and indirectly) challenges. Since it takes at most $pd$ seconds (see Section \ref{sec:distrib}) for a challenge to indirectly propagate across a network, a reasonable time interval $s$ for solving puzzles would be at least $pd$ seconds. Note that $s$ does have a lower bound in that the slowest user in the system has to be able to solve at least one puzzle in $s$ seconds. 

Whenever a node solves a puzzle, it records the {\em puzzle state}, which is data to verify the correctness and freshness of the puzzle solution. This includes the puzzle solution, the puzzle challenge, and that challenge's state record. To provide puzzle verification, nodes provide their latest puzzle state. Note that only the latest puzzle state is necessary for correctness because even nodes across the diameter of the network will have on record their challenges that indirectly were used to create the challenge used for the puzzle solution (see Section \ref{sec:distrib}). 

Say node $A$ contacted its neighbor $B$ during a system operation, such as a DHT lookup.  In addition to replying to the lookup query, $B$ responds with its latest puzzle state. The verifying node $A$ can validate the freshness of the puzzle solution by confirming that its own challenge stored in the received state record is still on record. This guarantees that $B$'s puzzle was solved within the last $2s+p(d+2)$ seconds. Using the remaining puzzle state data, $A$ can compute and use the puzzle challenge to authenticate $B$'s puzzle solution. If any step in the validation fails, $A$ can assume $B$ is misbehaving and refuse $B$'s service, inhibiting $B$'s negative influence.

\subsubsection{Verifying Non-Neighbors along a Path}
\label{sec:pathVer}

Neighbors-only challenge distribution prevents direct non-neighbor verification. However, non-neighbors can still be indirectly verified through {\em multi-hop verification}. During multi-hop verification, a node iteratively verifies other nodes along a path through a neighbor. This path could be a DHT lookup or routing path. The process of verifying puzzles at a specific node in the path is the same as described in Section \ref{sec:puzVer}, however, additional information is communicated and used during path verification.

As described in Section \ref{sec:puzVer}, when node $A$ verifies node $B$, $B$ responds with its puzzle state. However, when verifying a path, $B$ additionally sends its recorded challenges and their associated state tables. This allows $A$ to verify the correctness of any one of $B$'s recorded challenges by checking that the challenge from $A$ stored in the associated state table is still on $A$'s record, and recomputing the challenge to ensure it matches $B$'s challenge. Using this method, $A$ should be able to verify the correctness of most, although possibly not the tail-end (oldest challenges), of $B$'s record.

$A$ temporarily stores the verified portion of $B$'s record for the next verification. The following node $C$ in the path should be a neighbor of $B$, and $C$'s puzzles and challenges should be computed based on challenges from $B$. $A$ can now contact and verify $C$'s puzzle state and challenge records using the temporarily stored $B$ records, instead of $A$'s own records. If $C$ is verified, $A$ can again verify a portion of $C$'s record, then replaces its temporary records for $B$ with the verified portion of $C$'s records. In this manner, $A$ can iteratively continue to verify all nodes along a path, providing indirect verification of any node in the network.

\subsection{Protection Mechanisms}
\label{sec:protect}

Section \ref{sec:collect} describes a scheme that creates an environment where nodes can verify computational work of other nodes. This section discusses mechanisms that protect honest nodes from the forms of Sybil attack that can occur in that environment. There are three general strategies for a Sybil attack. 

The first strategy for a Sybil attack is to use consistently unverifiable Sybil nodes. These Sybils never perform computational work, allowing an adversary to support a large scale attack. Section \ref{sec:sync} details complete protection from this attack strategy, regardless of the scale of attack. The second attack method is using initially verifiable Sybil nodes, which bypass the defense in Section \ref{sec:sync}. After incorporation into the network, these Sybils halt computational work so an adversary can attempt to join and support even more malicious nodes. Section \ref{sec:tableVer} provides simple mechanisms that control the influence from such an attack. The mechanisms described in these two sections inhibit the influence of unverifiable nodes, which would force adversaries to use consistently verifiable Sybils, the final form of a Sybil attack. These Sybils correctly follow the collective verification scheme, making them indistinguishable from honest nodes, but may misbehave otherwise. However, an adversary with limited resources can only support a restricted number of verifiable Sybils, reducing their malicious influence. Section \ref{sec:virt} provides a mechanism that leverages heterogeneity in node computational capacities for further minimizing the impact of this attack form.

\subsubsection{Protecting against Consistently Unverifiable Sybils}
\label{sec:sync}

With the collective verification scheme, nodes can distinguish those who perform computational work from those who do not, allowing controlled communication with only behaving nodes. However, some system operations sensitive to network dynamics, particularly churn-handling protocols, may benefit from temporarily accepting unverified nodes for faster convergence. Maintaining this benefit may permit a large number of consistently unverifiable Sybils to hijack neighbor relationships through churn-handling.

To remedy this adversarial situation, SybilControl-enable nodes use a {\em delayed-adding} mechanism, where nodes only establish neighbor relationships upon computational work verification. Thus, a Sybil that never solves a puzzle will never be considered a neighbor of any nodes, and will have no influence. With the delayed-adding mechanism, nodes newly discovered through churn-handling are temporarily managed in an {\em untrusted list}. Challenges are distributed to both established neighbors and nodes in the untrusted list. Once an untrusted node should become verifiable, a direct verification is conducted, and only upon success does that node become an established neighbor.

This mechanism provides complete protection from consistently unverifiable Sybil nodes in that those nodes will never hijack neighbor relationships. Furthermore, newly discovered honest nodes are not immediately verifiable anyway, and should not be considered viable neighbors.  However, churn-handling protocols often rely on distributing information about neighbor relationships, meaning newly discovered nodes would not be included in churn-handling until verification. This can degrade churn-handling performance, which may not be acceptable to systems sensitive to churn. 

To maintain efficient protocol performance, churn-handling protocols include nodes in the untrusted list. If churn-handling protocols are utilizing information about the $i$-th neighbor, and there is a node in the untrusted list that should soon replace that neighbor relationship, then the protocol distributes information on the untrusted node. This allows churn-handling protocols to maintain convergence times by operating as if the newly discovered node was immediately established as a neighbor. Note that this does not reduce the protection of the delayed-adding mechanisms, as a consistently unverifiable node will remain only in the untrusted lists of nodes during churn-handling. Therefore, consistently unverifiable Sybil nodes simply cannot obtain positions to maliciously influence the system.

\subsubsection{Protecting against Initially Verifiable Sybils}
\label{sec:tableVer}

The mechanism in section \ref{sec:sync} removes the threat of consistently unverifiable Sybils. However, a malicious node can bypass the defense by initially performing work until it is no longer in the untrusted list, and has become an established neighbor. The adversary, now no longer paying a computational cost, may attempt to join more nodes. This vulnerability arises because nodes are validated in the collective verification scheme only upon contact, which may be infrequent.

To provide nodes with protection from this attack, two mechanisms are introduced. The first mechanism is the \textit{backup neighbors list}. When churn-handling with the delayed-adding mechanisms replaces an old neighbor relationship with a new neighbor, the old neighbor is still maintained in the backup neighbors list if it is currently \textit{safe}. A node can be determined safe by simply conducting a direct verification. This backup list serves as a level of redundancy should the primary neighbors become overrun with initially verifiable Sybils. Eventually, these Sybil will be detected and removed from the primary neighbors, which can be repopulated using both churn-handling and backup neighbors. Note that challenges must still be sent to nodes in the backup table because they may be contacted should a primary neighbor fail.

To ensure that a node's neighbors do not go unverified for long periods of time, the second defense mechanism is simple: \textit{periodic neighbor verification}. Neighbor and backup neighbors can be periodically verified, much like neighbor relationships are periodically updated with churn-handling protocols. If any neighbors fail verification, they can be replaced with a new node through using churn-handling or the backup list. Thus, adversaries can enter only a limited number of initially verifiable Sybils, since Sybils no longer performing work will be soon removed from all neighbor relationships and become uninfluential.

Ideally, neighbor relationships would be verified every puzzle time $s$, to ensure all neighbors are consistently performing work. However, the main purpose of periodic neighbor verification is to ensure that at least a subset of a node's neighbors is still fully verifiable and can be used for system operations. Then, verifying a fraction of neighbors every $s$ seconds can still be effective if the portion is large enough such that the node can still function if the non-verified portion was composed of unverified nodes.  We leave it to specific applications to determine the most appropriate frequency of this periodic table verification.

If an adversary has large amounts of computational resources, it could support enough initially verifiable Sybils to fill up both the primary and backup neighbor list of some nodes. SybilControl cannot prevent this from occurring and does not offer guaranteed protection. However, use of additional backup lists can provide more protection. Furthermore, Section \ref{sec:virt} discusses a mechanism that can reduce the initial influence of these initially verifiable Sybils, which may further reduce this threat.

\subsubsection{Leveraging Heterogeneity to Protect Against Consistently Verifiable Sybils}
\label{sec:virt}

The final strategy of a Sybil attacker is to attempt to support consistently verifiable Sybil nodes. These nodes are indistinguishable from honest nodes because they follow the collective verification scheme correctly, but may misbehave otherwise. While SybilControl cannot prevent these ``invisible'' Sybils from doing any harm, it can provide a mechanism that reduces the total influence these Sybils may have. The mechanism is \textit{virtual node usage}, which leverages heterogeneity in node computational capacities. 

Computational puzzle schemes tend to suffer from the heterogeneity of node capacities in a system. However, this heterogeneity is leveraged by SybilControl for improved Sybil protection. Puzzles must require a maximum amount of computational resources that should be manageable by nodes with less capacity. However, many honest nodes will have significantly more computational power than that maximum. Virtual node usage is a mechanism where honest users are allowed to optionally control extra virtual nodes if they have the resources to support them.  With more honest nodes, a larger fraction of the network is controlled by honest users, and Sybil attacks would have less influence. Note that it is inconsequential that we cannot distinguish honest from malicious users when assigning virtual nodes because a malicious user would support extra Sybil nodes regardless.

To provide some intuition as to the benefits of using virtual nodes, assume we have an $n$ node DHT network. On average, each node controls $\frac{1}{n}$ of the keyspace. If an adversary can join a maximum of $m$ Sybil nodes into the system, the adversary can control on average $\frac{m}{n+m}$ of the keyspace. If the average honest user has enough computational resources to maintain work for $q$ nodes, then by allowing users to support $q$ virtual nodes, the $m$-node Sybil attack will now only influence an average of $\frac{m}{q*n+m}$ of the keyspace, which can be a significantly smaller portion. This illustration shows that use of virtual nodes can protect more of the keyspace and reduce the influences of Sybil nodes that an adversary can support. In addition, the number of these Sybils should already be limited by the collective verification scheme and the mechanisms in Section \ref{sec:sync} and \ref{sec:tableVer}, which prevent unverified nodes from being a significant threat.

\section{A SybilControl-Enabled DHT}
\label{sec:incorp}

SybilControl creates a unique environment for distributed systems since nodes may or may not be verifiable. In this section, we discuss practical issues associated with the deployment of SybilControl. 
To provide concrete examples of techniques, we describe SybilControl in the context of DHTs. Many distributed systems, such as distributed file systems, distributed databases, and peer-to-peer networks, rely on a DHT as an underlying structure for storage or routing. While we describe this section in the context of DHTs for clarity, the general principles should be applicable to most other distributed systems. 

In Section \ref{sec:problem}, we discuss several practical issues introduced by deploying SybilControl. The techniques for a practical SybilControl-enabled DHT involve resilient lookup protocols described in Section \ref{sec:lookup}, and replication mechanisms as described in Section \ref{sec:rep}.

\subsection{Issues with Deploying SybilControl}
\label{sec:problem}

In a SybilControl-enabled system, nodes may not immediately trust communications with certain other nodes, such as newly joined nodes or un-established neighbors. SybilControl uses the delayed-adding mechanism, detailed in Section \ref{sec:sync}, to handle these untrusted nodes and include them in churn-handling protocols. While information about these changing neighbor relationships are propagated throughout the network, they do not manifested into new viable neighbors until the untrusted nodes provide proof of computational work, which requires some time for puzzle completion.  This delay between when a new relationship is discovered and when it becomes viable can result in increased sensitivity to churn in SybilControl-enabled systems.

For example, in a DHT such as Chord, a newly joined node $A$ immediately gains control over a portion of keyspace that previously was supported by its immediate successor $B$. In a SybilControl-enabled DHT, the delayed-adding mechanism of $B$ prevents it from immediately transferring control to $A$ as it has not been established as a verifiable neighbor yet. This is desirable for Sybil defense since the newly joined node could be an unverified Sybil. However, $B$ could fail before $A$ becomes verifiable, and all data $B$ may be lost, including those that would have been transferred to $A$. This possible scenario, among many others, is created by the SybilControl environment, and must be handled to avoid system performance degradation.

\subsection{Resilient Lookup Protocols}
\label{sec:lookup}

The primary DHT operation is the lookup for a key, where a node determines the path to the objects or nodes associated with that key. To prevent lookup performance degradation,  a SybilControl-enabled DHT can use resilient lookup protocols. These resilient lookups not only improve lookup performance with SybilControl, but are generally used to provide resilience against Sybil attacks. Many DHTs already support resilient lookups, so use of this technique does not require extensive modifications to existing DHTs.

Since SybilControl requires iterative path verification, multi-hop lookups in SybilControl systems must be iterative. To improve lookup resilience, backtracking and redundant lookups can be used. With backtracking lookups, the lookup initiator maintains a history of contacted nodes, and backtrack to the last contacted verifiable node when encountering a failed or unverified node. This results in lookups resilient to node failures or unresponsive Sybils. Lookup query messages can additionally carry the list of contacted nodes, including failed nodes. This provides a heuristic for nodes responding to lookup queries to ignore already-contacted nodes when determining the next hop in the lookup path. DHT lookups are designed to provide the best next-hop suggestion, and revisiting nodes would result in less-than-optimal cycles. 

With redundant lookups, multiple backtracking lookups should be conducted for the same key while sharing a common list of contacted nodes. This can result in multiple unique lookup paths. In the face of Sybil attacks or failed nodes, redundant lookups are more resilient because they fail only if all lookups contact Sybil or failed nodes, which is less likely with a larger number of redundant lookups.

\subsection{Replications}
\label{sec:rep}

In DHTs, the failure of a node maintaining some data can result in the loss of that data. A standard idea to improving DHT robustness, which applies also when enabling SybilControl, is to replicate, by storing additional copies of that data. With object-storage DHTs, where nodes maintain some data or files such as in a distributed file system, several replications can be supported by different nodes. Even if one node fails or is misbehaving, several other replicas are accessible. In routing DHTs, nodes hold pointers or addresses to other nodes to be used for routing. With replications, several nodes maintain the same pointers or addresses, instead of only one. Similar work has been done with the i3 Internet Indirection Infrastructure system \cite{i3}.

Replications can be maintained locally with respect to identifiers. With local replications, a node and several consecutive predecessors and/or successors can maintain redundant copies. It is less likely that these nodes will all simultaneously fail or be subverted by an adversary. However, the system is most robust if non-local nodes also support redundant copies. This prevents a targeted attack or a local network failure to a specific area of the identifier space to result in lost access to all data. The details of this are more specific to the underlying system, and are beyond the scope of our discussion.

 \begin{figure*}[th]
        \centering
        \subfigure[]{\includegraphics[width=0.48\textwidth]{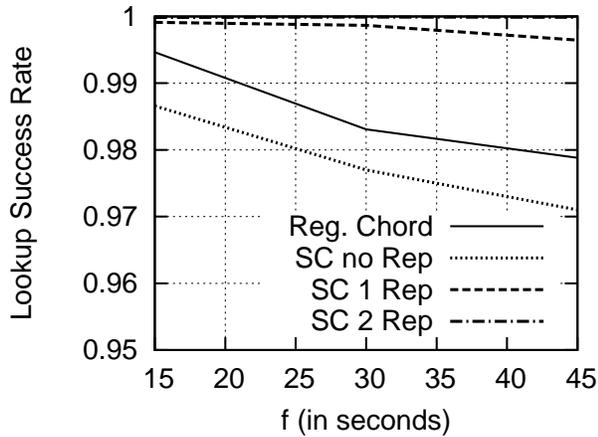} \label{fig:noSybilRate}}
        \subfigure[]{\includegraphics[width=0.48\textwidth]{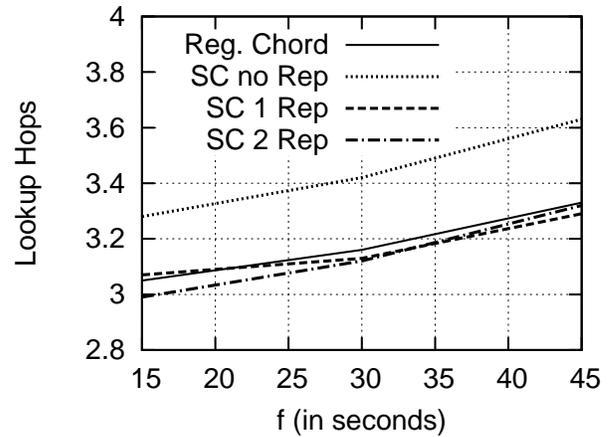} \label{fig:noSybilHop}}     
        \caption{\label{fig:noSybil} Lookup performance of regular Chord and SybilControl using different number of replications versus $f$, the frequency of churn-handling.  (a) shows average lookup success rates, while (b) graphs the lookup hops.}  
 \end{figure*}

\begin{figure*}[th]
        \centering
        \subfigure[]{\includegraphics[width=0.48\textwidth]{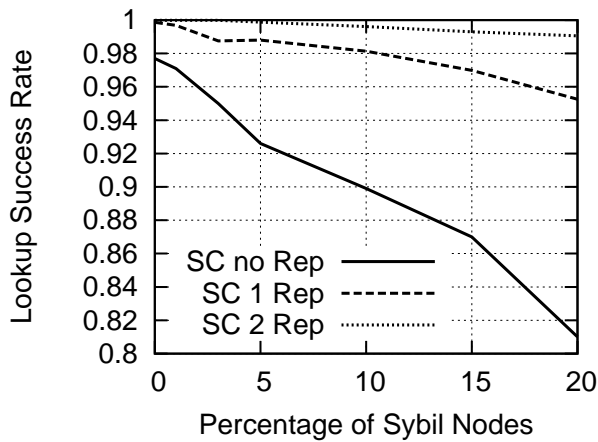} \label{fig:sybilRate}}
        \subfigure[]{\includegraphics[width=0.48\textwidth]{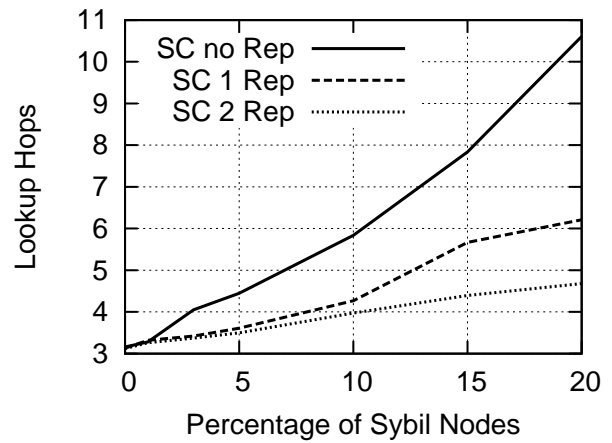} \label{fig:sybilHop}}     
        \caption{\label{fig:sybil} Lookup performance of SybilControl using different number of replications when varying percentages of Sybil nodes in the network.  (a) shows average lookup success rates, while (b) graphs the lookup hops.}  
 \end{figure*}
 
\section{Evaluation}
\label{sec:eval}

In this section, we evaluate the performance overhead of enabling SybilControl in a DHT and the ability of SybilControl to protect the DHT from attack. We investigated these properties using our own simulator of the Chord DHT. We chose Chord for its simplicity and scalability. The simulator directly implements standard Chord protocols as well as the protocols and mechanisms described in Section \ref{sec:design} and Section \ref{sec:incorp}. 

To allow for consistent comparisons, the simulations followed the same network dynamics. Initially, the network consisted of 1000 verifiable nodes. Nodes sent ping messages carrying challenges every 5 seconds. We use a puzzle time of 20 seconds, which is based off our suggestion for a puzzle time larger than $pd$ in Section \ref{sec:distrib}. We model churn similarly to other studies\cite{churn}, with a pareto distribution for mean session time and an exponential distribution for mean downtime. Since the same studies indicate that the mean session time is typically more than 60 minutes, we used 60 minutes as our distribution means. 

Chord uses two churn-handling protocols: fix-fingers and stabilize. In the original Chord evaluation \cite{chord}, the frequency $f$ of each protocol execution is in a range between every 15 seconds to every 45 seconds. We use values of $f$ also within that range. Each simulation is ran for 10000 seconds before measurements, to avoid cold-start effects.

\begin{figure*}[t]
        \centering
        \subfigure[No Replication]{\includegraphics[width=0.32\textwidth]{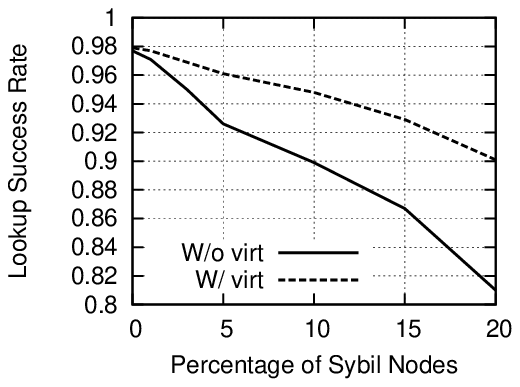} \label{fig:virt0}}
        \subfigure[One Replication]{\includegraphics[width=0.32\textwidth]{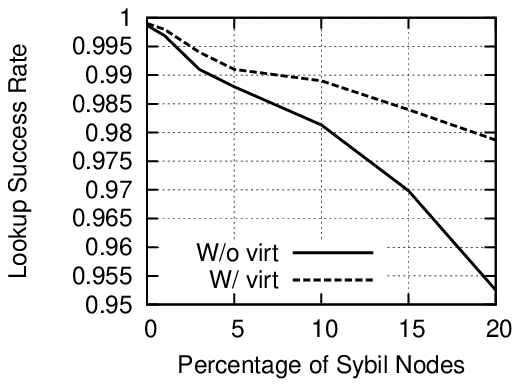} \label{fig:virt1}}  
        \subfigure[Two Replications]{\includegraphics[width=0.32\textwidth]{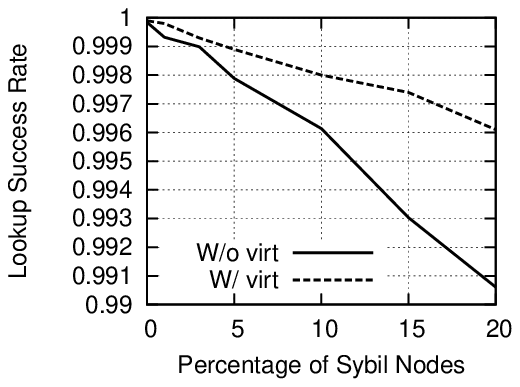} \label{fig:virt2}}   
        \caption{\label{fig:virt} Lookup success rates of SybilControl without virtual nodes and SybilControl where honest users support an average of 2 virtual nodes, while varying percentages of the initial network are Sybil nodes. }  
 \end{figure*}
 
 \begin{figure*}[t]
        \centering
        \subfigure[No Replication]{\includegraphics[width=0.32\textwidth]{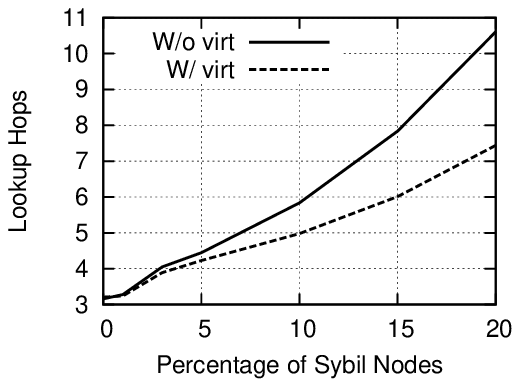} \label{fig:virt0Hop}}
        \subfigure[One Replication]{\includegraphics[width=0.32\textwidth]{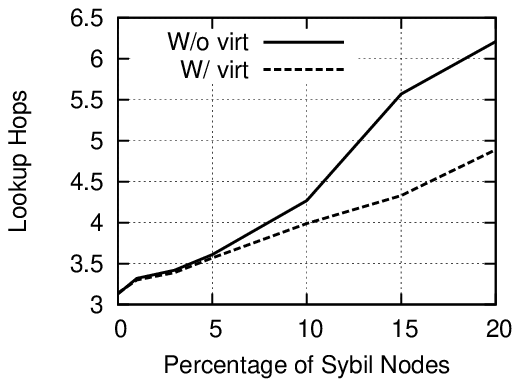} \label{fig:virt1Hop}}  
        \subfigure[Two Replications]{\includegraphics[width=0.32\textwidth]{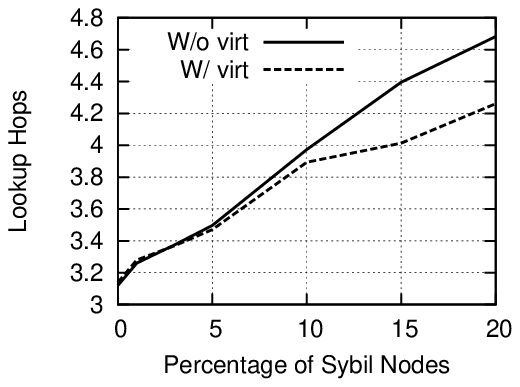} \label{fig:virt2Hop}}   
        \caption{\label{fig:virtHop} Lookup hop counts of SybilControl without virtual nodes and SybilControl where honest users support an average of 2 virtual nodes, while varying percentages of the initial network are Sybil nodes.}  
 \end{figure*}
\subsection{Performance Overhead of SybilControl Deployment}

This section investigates the performance overhead of enabling SybilControl 
in the Chord DHT. The primary operation in Chord is a lookup, so we evaluated 
lookup performance metrics for Chord with and without SybilControl enabled. 
Each alive node in the network attempted 10 resilient lookups for random keys. 
We measured the success rate of attempted lookups and the number of hops in a 
successful lookup's path, which we define as the number of nodes contacted, 
including failed nodes. To accurately measure only performance overhead, 
the network excluded any misbehaving nodes. $f$, Chord's fix-finger and 
stabilize time interval, is varied between every 15, 30, and 45 seconds.
 Since a smaller $f$ implies more frequent churn-handling, we expect that 
$f$ is negatively correlated with lookup performance.

Since SybilControl systems must additionally deal with node verifiability, 
it would not be surprising if SybilControl reduced performance. However, 
the  techniques in Section \ref{sec:incorp} should address these issues and 
improve performance. In particular, using multiple replication in SybilControl 
is expected to help improve performance. With more replicas, SybilControl 
lookups should be more likely to succeed in finding an alive, verifiable 
node that holds a copy. SybilControl was evaluated using 0, 1, and 2 replication.

Figure \ref{fig:noSybil} shows the lookup performance measurements for 
SybilControl-enabled Chord using varying number of replications and 
regular Chord. Figure \ref{fig:noSybilRate} plots the success rate of 
lookups versus $f$. As expected, regular Chord outperformed SybilControl 
without replications, with lookup success rate about 1$\%$ higher for regular Chord. 
However, even when one extra replication was used, SybilControl's performance improves 
significantly and is better than regular Chord. At two replications, average lookup 
success rate is consistently near 100$\%$. The effect of varying $f$ also generally 
matched expectation. Lookup success rates for regular Chord and no-replication 
SybilControl decline with larger $f$, by as much as 2$\%$. One-replication SybilControl 
seems to also be affected to a lesser degree. Performance with two replications seems unaffected by $f$.

Figure \ref{fig:noSybilHop} plots the lookup hops versus $f$. Again, 
no-replication SybilControl performs worse than regular Chord, typically 
by approximately 0.3 hops. However, once one replication is introduced, 
SybilControl lookups roughly match regular Chord lookups in hop count, 
varying by less than 0.1 hops. Additional replications only give diminishing 
improvements. The effect of $f$ is more apparent in this graph. Increase in $f$ 
results in an increase in the average number of contacted nodes for both regular 
Chord and SybilControl. Increased numbers of replications do not seem to show a 
resistance to differences in $f$ as they do with lookup success rate.

\label{overhead-eval}
\subsection{SybilControl-Enabled DHT Under Attack}

In this section, we look into performance of a SybilControl-enabled DHT when an adversary is performing a Sybil attack. SybilControl limits the number of Sybil nodes an adversary can support but it does not provide strict bounds on that number. For our evaluations, we assume the adversary has enough computational resources to support up to 20$\%$ of the nodes in the initial 1000 node network. Note that this may represent scenarios worse than what is likely to occur in a real-world SybilControl systems.

We use the same performance metrics as described in Section \ref{overhead-eval}. $f$ is set at every 30 seconds for all simulations. For this evaluation, we assume Sybil nodes drop any lookup queries. Simulating more complex behavior would wander into the realm of other attacks, which may be dealt with in other works.  For resilient lookup protocols, nodes first perform one lookup with a time-to-live of 30 hops, and upon failure, perform two simultaneous redundant lookups. Also, we do not draw comparisons with regular Chord because it was not designed to withstand Sybil attacks. Again, we tested SybilControl using 0, 1, and 2 replications. To study the effect of including virtual nodes, we also tested SybilControl where each honest user could support an average of 2 virtual nodes.

With additional replications, we expected that Sybil resistance would improve. More replications increase the chance that an honest, verifiable node hold a replica. Also, use of virtual nodes is expected to also improve Sybil resistance. With more virtual nodes, each Sybil node should claim a smaller portion of the key space, limiting the total influence of an adversary.

\subsubsection{Results using Replications}

Figure \ref{fig:sybil} graphs the lookup performance metric for SybilControl using different number of replications, when under varying strength Sybil attacks. Figure \ref{fig:sybilRate} plots the SybilControl lookup success rate versus percentage of Sybil nodes in the network. As expected, the use of replication is beneficial. No-replication SybilControl was significantly affected by Sybil attacks, with lookup success rate almost 20$\%$ lower when the Sybil nodes made up 20$\%$ of the network. Using even a single replication dramatically reduced this drop-off. Single replication saw a lookup success rate drop-off of about 4$\%$ under the same powered attack. SybilControl using two replications appears to be almost unaffected by Sybil nodes, retaining above 99$\%$ success rate.

Figure \ref{fig:sybilHop} plots the lookup hops versus percentage of Sybil nodes in the network. Again, no-replication SybilControl performed poorly, resulting in more than a three-fold increase in lookup hops when under large Sybil attacks. Use of replication significantly reduces this increase, but is not completely resistant against Sybils. Single replication SybilControl under a Sybil attack composing of 20$\%$ of the network still experiences a two-fold increase in lookup hops. While the benefit of each additional replication still diminishes, it does not appear to diminish as fast as when free of Sybil influence. Again under the 20$\%$ network Sybil attack, switching from one to two replications decreased lookup hops by 1.3 nodes, while switching from no to one replication lowered the lookup hops by 4.1 nodes. Using even more than two replications may lead to further increases in Sybil resistance.

\subsubsection{Results using Virtual Nodes}

Figure \ref{fig:virt} depicts the lookup success rate of SybilControl systems under Sybil attack, where the average honest user supports 2 virtual nodes. In each graph, a different number of replications were used. As predicted, use of virtual nodes improved lookup success in general. However, virtual nodes were most effective when using fewer replications. When Sybils controlled 20$\%$ of the initial network (before virtual node usage), use of virtual nodes improved no-replication SybilControl lookup success rate by about 10$\%$, while two-replication SybilControl only improved by less than 1$\%$. This is expect since using large number of replications already improves success rate to near 100$\%$, so there is little room for improvements from virtual nodes.

Figure \ref{fig:virtHop} depicts the number of nodes contacted during SybilControl lookups using virtual nodes and under Sybil attack. As seen with lookup success rate, virtual nodes help decrease the number of nodes contacted, and the benefits are most apparent with fewer replications. With no-replication SybilControl, virtual node usage decreases the lookup hop count by about 3 nodes when 20$\%$ of the initial network is Sybils. Under the same attack, two-replication SybilControl saw only a decrease of 0.4 in the average number of contacted nodes, which is approximately a 10$\%$ improvement. An interesting observation is that virtual nodes do not seem to benefit much at all under smaller Sybil attacks, where the number of Sybils is less than 5$\%$ of the initial network. This isn't unexpected though, given that the use of virtual nodes decreases the influence of each Sybil node slightly, and with few Sybil nodes, the total improvement is less noticeable.

\section{Conclusion}
\label{sec:conclusion}

This paper presented SybilControl, a novel, decentralized scheme for controlling the extent of Sybil attacks. SybilControl consists of a distributed protocol to allow nodes to collectively verify computational work of their neighbors and defense mechanisms to enforce that nodes conduct computational work to stay functional in the system. Thus, adversaries with finite amount of computational resources will be only able to support a limited number of Sybil nodes. To demonstrate the practicality of SybilControl, it was deployed in the context of DHTs with the aid of existing DHT functionalities. The performance overhead when enabling SybilControl was shown to be manageable. Furthermore, a SybilControl-enabled DHT was shown to be resilient against large-scale Sybil attacks, capable of maintaining above 99$\%$ lookup success rate using low communication overhead.

\footnotesize
\bibliography{paper}
\bibliographystyle{abbrv}

\end{document}